\documentclass[reprint, aps, prb, amsmath, showpacs, superscriptaddress, groupedaddress]{revtex4-1}
\usepackage{graphicx}
\usepackage{epsfig}
\usepackage[cp1251]{inputenc}
\usepackage[english]{babel}
\usepackage{amsmath}
\usepackage{amssymb}
\usepackage{amstext}
\usepackage{color}

\begin{document}

\title{Semiclassical analysis of intraband collective excitations
 in a two-dimensional electron gas with Dirac spectrum}
\author{S. M. Kukhtaruk, V. A. Kochelap}

\affiliation{Department of Theoretical Physics, V.E. Lashkaryov Institute of Semiconductor Physics  NASU,
 Pr. Nauki 41, Kiev 03028, Ukraine}

\begin{abstract}
{Solving the initial value problem for semiclassical equations that describe two-dimensional electrons with the Dirac spectrum
we found that collective excitations of the electrons are composed by a few distinct components of
the oscillations. There always exist sustained plasma oscillations
with well known plasmon frequency $\omega_{pl}(k)$. Additionally there
are oscillations with the 'carrier frequency' $\omega = v_Fk$
slowly decaying in time according to a power law  ($v_F$ and $k$
are the Fermi velocity  and wavevector). The reason for onset of these
oscillations has a fundamental character related to branching of the polarization
function of the Dirac electrons. A strongly anisotropic initial disturbance
of the electron distribution generates additional  component of undamped
oscillations in the form of an electron unidirectional beam which are van Kampen's modes in the Dirac plasma.
}
\end{abstract}
\maketitle

Polarization properties of electrons determine their response to
an external electrical signal, define collective plasmon and
plasmon-phonon modes, as well as many other phenomena involving
interaction of the charges. A fundamental and applied aspects of collective excitations in graphene-like systems were discussed in detail in several reviews~\cite{Castro-rev, SarmaReview,NovoselovReview,PlasmonReview, BasovReview, PlasmonReview1,PlasmonReview2,PlasmonReviewTHz}.
The polarizability of two-dimensional (2D) electrons with the linear Dirac
spectrum, $\textstyle{\epsilon ({p}) = v_F p}$, differs considerably from that
of the electrons with a parabolic spectrum ($\textstyle{\epsilon (p)\propto p^2}$)
characteristic for conventional low-dimensional heterostructures.
Particularly, the absence of the spectrum curvature leads to
less effective screening of the Coulomb potential \cite{Castro-rev},
gives rise to the divergence of the irreducible polarizability \cite{Wunsch, Sarma, Ryzhii_gate}
$\Pi (\omega, k) \propto 1/\sqrt{\omega^2 - v_F^2 k^2}$
at $\textstyle{\omega \rightarrow \pm v_F k}$  with
$\omega$ and ${\bf k}$ are frequency and the 2D wavevector,
respectively, $v_F$ is the Fermi velocity.
Then, the square-root behavior of the polarizability means that
$\Pi (\omega,  k)$ is a {\it double-valued function} on the
complex $\omega$-plane.  For calculation of real characteristics of
the physical system, properties of the polarizability as an analytical
function on the complex $\omega$-plane is critically important.
The semiclassical approach based on the analysis of the
Boltzmann-Vlasov system of equations allows one to construct the \emph{principal branch}
of the polarizability function $\Pi (\omega, k)$ and obtain transparent
and easy interpreted results. These results \emph{complement} those calculated with the
use of different quantum mechanical approaches in the long-wavelength
limit \cite{Wunsch, Sarma}.

Recently, new nanoscope techniques were
proposed to launch the excitations by a sharp tip of atom force microscope
and to monitor
them by the scattering type near-field optical microscope~\cite{NS-1,NS-2}.
The techniques have enabled the experimental exploration of spatio-temporal~\cite{Fei-1,Chen-1,Fei-2,Fei-3}
and time-resolved~\cite{Fei-4} dynamics of the collective excitations in the
graphene and graphene-like systems.
Importantly, both the excitation amplitude and phase
can be measured~\cite{PRL-phase}. Similar studies have been performed by the use of the resonant antenna plasmon launcher and near-field optical microscope~\cite{Alonso}. Another promising method for plasmon investigation is time-resolved electrical measurements~\cite{Japan-1, Japan-2}.
Observed in these works macroscopic effects - long-wavelength charge modes and local electric fields -
can be described by using the semiclassical approach.

In this Communication, we present the semiclassical analysis of
{\it longitudinal intraband}
excitations of 2D electron gas with the Dirac spectrum.
It is appropriate to note, that the transverse electric mode
in graphene-like systems was analyzed in paper~\cite{Mikhailov}.
We assume the $n$-doped system and restrict ourselves to
excitations with $\omega$ and ${\bf k}$ for which
interband processes can be neglected at a given electron concentration, $n$,
and ambient temperature, $T$.
Such approach is valid for $\textstyle{\hbar \omega < \epsilon(p_F),\, \hbar k < p_F}$
at $\textstyle{k_B T \ll \epsilon(p_F)}$, with $p_F$ being the Fermi momentum and $k_B$
is the Boltzmann constant
(detail discussion see, for example, in Ref.~\onlinecite{Castro-rev}).

The Boltzmann-Vlasov system of equations consists of the collisionless transport equation
\begin{equation} \label{BTE-1}
\frac{\partial F}{\partial t} + v_F \frac{\bf p}{p} \left. \frac{\partial F}{\partial {\bf r}}
+e \frac{\partial\Phi}{\partial \mathbf{r}}\right|_{z=0} \frac{\partial F}{\partial {\bf p}} = 0,
\end{equation}
and the Poisson equation
\begin{equation} \label{Poisson}
\Delta \Phi =  \frac{4 \pi e \delta[z]}{\kappa}
\int \frac{g\,d^2 p}{(2 \pi \hbar)^2} [F({\bf r, p },t) - F_0({\bf p }) ]\,.
\end{equation}
Here the Dirac spectrum is assumed for the electrons confined to the sheet at $\textstyle{z=0}$.
The electron coordinate and momentum
$\bf r,\,p$ are the 2D vectors in the $x-y$ plane.
$F({\bf r,p},t)$ is the electron distribution function, $F_0 ({\bf p})$
is that under equilibrium. $\Phi ({\bf r},z,t)$ is the self-consistent electrostatic potential.
In Eq.~(\ref{Poisson}), $e$ is the elementary charge,
$\delta [z]$ is the Dirac delta-function, $g$ is the degeneracy factor
of the electron band (for graphene $g = 4$).  The parameter
$\kappa$ depends on dielectric environment. If the electron sheet is between
two materials with the dielectric constants $\kappa_l,\,\kappa_h$, in final
formulae one shall use $\textstyle{\kappa = (\kappa_l+\kappa_h)/2}$.

For the collisionless limit it is necessary $\textstyle{\omega \tau \gg 1}$, $\textstyle{v_Fk \tau \gg1}$, with
$\tau$ being a characteristic scattering time. For our purposes the equation \eqref{BTE-1} shall be linearized.
We denote variation of the distribution function as
${\cal F}({\bf r},{\bf p},t)\equiv F({\bf r},{\bf p},t) - F_0({\bf p}) $
and set ${\cal F}({\bf r},{\bf p},t) = {\cal F}_{\bf k} ({\bf p},t)\,exp(i {\bf k r})$,
and ${\Phi}({\bf r},z,t) = {\Phi}_{\bf k}(z,t)\,exp(i {\bf k r})$.
Now Eqs.~(\ref{BTE-1}), and (\ref{Poisson}) take the form:
\begin{equation} \label{BTE-2}
\frac{\partial {\cal F}_{\bf k}}{\partial t} + i v_F \frac{\bf k p}{p} {\cal F}_{\bf k} = - ie {\bf k}{\Phi}_{\bf k} (0,t) \frac{\partial F_0}{\partial {\bf p}}\,,
\end{equation}
\begin{equation} \label{Poisson-2}
\frac{ d^2 {\Phi}_{\bf k}}{d z^2} - k^2 \Phi_{\bf k} =  \frac{4 \pi e \delta[z] }{\kappa}
\int \frac{g\,d^2 p}{(2 \pi \hbar)^2} {\cal F}_{\bf k}( {\bf p},t) \,.
\end{equation}
Following the Landau approach \cite{Landau-1}, we consider the initial
value problem by using the Laplace transform:
\begin{equation} \label{Laplace-1}
f_{\omega,\bf k}({\bf p})\! =\!\!\! \int\limits^{\infty}_0 \!\! {\cal F}_{\bf k} ({\bf p},t)\,e^{i \omega t} dt ,\,\,{\cal F}_{\bf k} ({\bf p},t)\! \!\!=\!\!\!\int\limits^{\infty + i \sigma}_{-\infty +i \sigma}\!\!\!\!\!\!\! f_{\omega,\bf k}({\bf p}) \,e^{-i \omega t}\,\frac{d \omega}{2 \pi}\,,
\end{equation}
where $\sigma >0 $. Similarly, we define the transformation of the potential, $\phi_{\omega, \bf k}(z)$.
Now one can easily find the solution for $f_{\omega,\bf k}({\bf p})$:
\begin{equation} \label{sol-f}
f_{\omega,\bf k}({\bf p}) = i \frac{{\delta\cal F}_{\bf k}({\bf p})-
i e \phi_{\omega, \bf k}(0) {\bf k} \,d F_0({\bf p})/d{\bf p}}{\omega -
v_F ({\bf k p})/p}\,,
\end{equation}
with $\delta {\cal F}_{\bf k}({\bf p})$ being a given initial perturbation of the electron
distribution. The solution to Eq.~(\ref{Poisson-2}), which decaying at
$z \rightarrow \pm \infty$, also can be easily found. The potential at $z=0$ is:
\begin{equation}  \label{sol-phi}
\phi_{\omega, \bf k}(0)\! =\!\frac{ -  2 \pi i e }{ \kappa \,k \, \Delta(\omega, k)}
\!\!\int \!\! \frac{g \,d^2 p}{(2 \pi \hbar)^2}\!
\frac{\delta {\cal F}_{\bf k}({\bf p})}{\left[\omega- v_F ({\bf k p})/p \right]}\!\! \equiv
 \!\!\frac{N(\omega,{\bf k})}{ \Delta(\omega, k)}\,,
\end{equation}
with $\Delta (\omega, k) = 1 - 2 \pi e^2\,  \Pi (\omega, k)/ {\kappa\,k}$ and
\begin{equation} \label{delta-1}
 \Pi (\omega, k)\ \equiv - \int \!\!\! \frac{g \,d^2 p}{(2 \pi \hbar)^2}
\frac{({\bf k p}) /p }{\left[ \omega- v_F ({\bf k p})/p \right]}  \frac{d F_0(p)}{d p}\,.
\end{equation}
It is clear that $\Pi (\omega, k)$ has the meaning of the polarizability
obtained in the semiclassical limit. $F_0 ({\mathbf{p}})$ is the Fermi distribution, then
\begin{equation} \label{Pi-1}
 \Pi (\omega, k) = \frac{\kappa \, K}{2 \pi e^2}
\left[ \frac{1}{2\pi} \int^{2 \pi}_0 d \alpha \frac{\omega}{\omega - v_F k \cos {\alpha}} -1 \right]\,,
\end{equation}
\begin{equation} \label{Kp}
K= \frac{e^2 g k_B T}{\kappa\hbar^2 v_F^2} \ln{\!\left[\exp{\!\left[\frac{E_F}{k_B T}+1\right]}\right]}\,,
\end{equation}
where $E_F$ is the chemical potential.

According to the definition of the Laplace transform (\ref{Laplace-1}),
in foregoing Eqs.~(\ref{sol-f})-(\ref{Pi-1}) $\omega =\omega^{\prime}+ i \omega^{\prime\prime}$ is the \emph{complex variable} belonging to the upper half-plane, $\omega^{\prime\prime} >0$. Under the latter condition,
the integral in Eq.~(\ref{Pi-1}) can be calculated as
\begin{equation} \label{int-1}
\Pi(\omega, k) =   \frac{\kappa \, K}{2 \pi e^2}
\left(\frac{\omega}{\sqrt{\omega^2- v^2_F k^2}} -1 \right)\,.
\end{equation}
The function (\ref{int-1}) is double-valued
with two \emph{branch points}, $\omega = \pm v_F k$, as discussed in introduction.

Consider an \emph{analytical continuation} of $\Pi(\omega, k)$ given by (\ref{Pi-1})
to the lower half of the $\omega$-plane. We start with the use of  (\ref{int-1}) valid for all complex $\omega$-plane except
the segment at the real axis $ -v_F k \leq \omega^{\prime} \leq v_F k$.
By choosing the cut along this segment, we select the principle branch of the function
(\ref{int-1}). Defined by such a way, $\Pi(\omega, k)$ is continuous
when crossing the real axis at $|\omega| > v_F k$, while it undergoes a
jump in its value at the cut.

As a result of this procedure, we obtain also the function
$\Delta(\omega, k)$ as well-defined function
in the complex  $\omega$-plane. It is easy to see that this function
has simple zeros on the  real axis
\begin{equation}  \label{zeros-1}
\omega^{\prime}= \pm \omega_{pl}\,,\,\,\,\omega_{pl} (k) =  \frac{v_Fk \left( 1+ {k/K} \right)}
{\sqrt{\left( 1+ {k/K}\right)^2-1}}\,,
 \end{equation}
which are the exact solutions of the equation $\textstyle{\Delta(\omega, k)=0}$.
For what follows, it is important that $\omega_{pl} (k) > v_Fk$.

Before to proceed with the farther analysis, let us estimate the value
$K$ for the graphene. Assuming the graphene sheet over a $SiC$-substrate
($\kappa_l \approx 9.7,\,\kappa_h=1$), setting the electron concentration
$n =2 \times 10^{12} cm^{-2}$ and temperature $\textstyle{T=300}$ K, we obtain $E_F \approx 0.16\,eV$, $\textstyle{E_F/k_BT\approx 6.12}$, $p_F/\hbar \equiv k_F \approx 2.5 \times 10^6\,cm^{-1}$, and $\textstyle{K \approx 3.9 \times 10^6\,cm^{-1}}$. That is $k_F < K$ and for the semiclassical analysis we should use $\textstyle{k < k_F < K}$. For example, at $\textstyle{k = 0.05\,K}$
we obtain $\textstyle{\omega_{pl} \approx 6.4 \times 10^{13}\,s^{-1}}$ and $\textstyle{v_Fk \approx 2 \times 10^{13}\,s^{-1}}$. We refer to this set of parameters as $\kappa$-environment-I. To estimate the criteria of validity of the collisionless approximation, one
needs to know the characteristic scattering time, $\tau$.
The 'intrinsic' electron-electron scattering time is of the order of
$10^{-11}\,s$~\cite{Principi}.
Optical phonon scattering can be neglected for the above accepted parameters.
Acoustical phonon scattering time at $T \leq 300\,K$ is estimated to be less
than $2 \times 10^{-11}\,s$~\cite{SarmaReview}. Elastic scattering by imperfections
is dominant. It can be determined via the phenomenological relationship
for the transport time: $\tau \approx \mu E_F/e v_F^2$ with $\mu$ being
the mobility. Assuming $\mu =2.5 \times 10^{4}\,cm^2/V\,s$,
we obtain $\tau \approx 4 \times 10^{-13}\,s$ and find that the
semiclassical approximation criteria are met: $\omega_{pl} \tau \approx 26 \gg 1$,
$v_Fk \tau \approx 8 \gg 1$.

For the graphene sheet in a high-$\kappa$ environment (for example, graphene
in a solvent~\cite{Ferro}), the value $K$
can be sufficiently less than $k_F$. Then the case $\textstyle{k \geq K}$ may be actual.
For example, at $\textstyle{\kappa_l \approx 9.7}$, $\textstyle{\kappa_h = 50}$,
and $k=K$ ($\kappa$-environment-II) for the same electron concentration, we obtain
$K \approx 7 \times 10^{5}\,cm^{-1}$ and $\omega_{pl} \approx 8 \times 10^{13}\,s^{-1},\,
v_Fk \approx 7 \times 10^{13}\,s^{-1}$ (note, the accepted $k$ is greater than for
the $\kappa$-environment-I). These estimates show that the excitations
considered are rather of the THz diapason.

Now, one can perform the inverse Laplace transform to find desirable
functions in the time-domain. Below, we will concentrate on the potential at $z=0$:
\begin{equation}  \label{phi-2}
\Phi_{\bf k} (t) \equiv \Phi_{\bf k} (0,t) = \int\limits^{\infty + i \sigma}_{-\infty +i \sigma}
\!\!\!\!\!\!\phi_{\omega,\bf k}(0)\, e^{-i \omega t} \,\frac{d \omega}{2 \pi}.
\end{equation}

According to Eq.~(\ref{sol-phi}), one can present  $\phi_{\omega,{\bf k}} (0)$ as
a fraction, where  the nominator, $N (\omega,{\bf k})$, is determined by the initial
perturbation, $\delta{\cal F}_{\bf k} ({\bf p})$. First, we will use the typical
assumption~\cite{Landau-1}  that the initial perturbation is such
that $N (\omega,{\bf k})$ has no poles in the
$\omega$-plane. Then, the analytical properties of both $\Delta(\omega, k)$
and whole integrand in (\ref{phi-2}) allow one to deform the integration
contour and calculate explicitly contributions of the
residues related to the zeros of $\Delta(\omega, {\bf k})$ and branch points:
\begin{gather}
\Phi_{\bf k} (t) = \Phi^{R}_{\bf k} (t) + \Phi^{\cal C}_{\bf k} (t)\,, \label{phi-0} \\
\Phi^R_{\bf k} (t) = - i \frac{\left[\omega_{pl}^2-v_F^2k^2 \right]^{3/2}}{v_F^2 k K} \times \nonumber \\
\!\!\!\!\!\left[\phantom{\frac{A}{B}}\right. \!\!\!\!\!\!\!N(\omega_{pl},{\bf k}) \exp{[-i \omega_{pl}\, t]} -
N(-\omega_{pl},{\bf k}) \exp{[i \omega_{pl} \,t]} \left.\phantom{\frac{A}{B}}\!\!\!\!\!\!\!\right],  \label{phi-3} \\
\Phi^{\cal C}_{\bf k} (t) = \oint_{\cal C} \phi_{\omega,\bf k}(0) \,
e^{-i \omega t}\,\frac{d \omega}{2 \pi}\,. \label{phi-4}
\end{gather}
In the integral (\ref{phi-4}), contour $\cal C$ encloses the cut, as shown in the inset of
Fig.~1(a).

In Eq.~(\ref{phi-0}), the first term oscillating with time describes
the undamped plasmons excited by the initial perturbation of the distribution
function. The latter specifies only the
magnitude of the plasmon excitations, while their frequencies
do not depend on the initial perturbation. They are determined by
the zeros of $\Delta(\omega, k)$ given by Eq.~(\ref{zeros-1}).
For $k \ll K$, one obtains 'the square
root law' \cite{Wunsch, Sarma},  $\omega_{pl} \approx v_F \sqrt{k K/2}$;
in the opposite limit, $\omega_{pl} \rightarrow v_F k$.

The last term in Eq.~(\ref{phi-0}) arises due to the existence
of \emph{branch points} in $\Delta (\omega,  k)$.
For electron systems with regular \emph{quadratic} energy dispersion,
there is \emph{no branching} and such contribution does not exist.
The time dependence of this term is defined by the initial
perturbation. Let us consider a few examples of different initial perturbations.

First, assume that the initial perturbation is
isotropic, $\delta {\cal F}_{\bf k} ({\bf p}) = \delta {\cal F}_k ( p)$,
i.e., for perturbed electron gas the average momentum and
velocity are zero. Then,
\begin{equation}  \label{N-1}
N(\omega,{k}) = \frac{ i \Phi_{k}^0}{\sqrt{\omega^2-v_F^2 k^2}}\,,\,\,
\Phi_{k}^0 =\! - \frac{e g}{\kappa \hbar^2 k} \!
\int\limits^{\infty}_0\!\! p dp \, \delta {\cal F}_k (p)\,,
\end{equation}
where the square root should be defined as principal branch on the $\omega$-plane with the above discussed cut.
Now, the contribution of the residues to the potential equals
\begin{equation} \label{N1-contr}
\frac{\Phi^R_{k} (t)}{\Phi_{k}^0} =
\frac{ 2 \left[\omega_{pl}^2 (k)-v_F^2 k^2 \right] }{v_F^2 k K }\,cos[\omega_{pl} (k)\, t]\,.
\end{equation}
The contour integral of Eq.~(\ref{phi-4}) can be  easily estimated
for two limiting cases:
\begin{equation}
\displaystyle{\frac{\Phi^{\cal C}_{k} (t)}{\Phi_{k}^0}  \approx}
\begin{cases}
\displaystyle{\frac{k}{K} J_1 (v_F k t)/v_F k t,  k \ll K,}\\
\displaystyle{J_0 (v_Fk t),  \phantom{empty!!!!} k \gg k\,,}
\end{cases}
\end{equation}
where $J_0(x),\,J_1 (x)$ are the Bessel functions.
At $t \gg 1/(v_F k)$, both expressions can be further simplified
\begin{equation}
\displaystyle{\!\!\!\!\frac{\Phi^{\cal C}_{k} (t)}{ \Phi_{k}^0}  \approx\! \sqrt{\frac{2}{\pi}} \!}
\begin{cases}
\displaystyle{\frac{k}{K}
 \frac{\cos[v_F k t -3 \pi/4]}{(v_F k t)^{3/2}}, \,\,\,\, }& k \ll K\,,\\
\displaystyle{\,\, \frac{\cos[ v_F k t \!-\!\pi/4 ]}{(v_F k t)^{1/2}},}& k \gg K\,.
\end{cases}
\end{equation}

Thus, the initial isotropic perturbation of the electron distribution
generates two different components of electrostatic potential oscillating in time and space. The first component is, obviously,  sustained
regular plasmon oscillations, excited by the initial perturbation. The
second component corresponds to oscillations with the 'carrier' frequency
$\omega = v_F k$. The oscillations decay in time according to a power law.
Mathematically, they arise due to the existence of the branch points of
polarization function (\ref{Pi-1}). Recovering the space-dependent factor $exp[i {\bf k r}]$, one can see that $\Phi_{\bf k}^{R}$ and $\Phi_{\bf k}^{\cal C}$ components correspond to {\it standing} waves. The main panels of Figs. 1(a) and (b) show the normalized
potential $\varphi_{k}^{\cal C} (t) = \Phi_{k}^{\cal C}/\Phi_{k}^0$
of these oscillations for two $\textstyle{\kappa}$-environments discussed above.
\begin{figure}
\begin{center}
\includegraphics[width=8.6 cm, height=8.6 cm,%
keepaspectratio, trim=0 40 0 0]{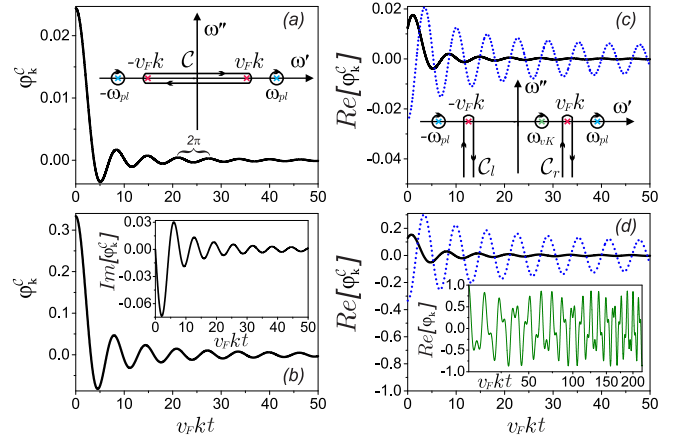}
\end{center}
\caption{(color online).
Normalized potential $\varphi_{\bf k}^{\cal C} (t)$ for: (a), (c) -- $\kappa$-environment-I and (b), (d) -- $\kappa$-environment-II;
(a), (b) -- isotropic initial perturbation;
(c), (d) -- strongly anisotropic initial perturbation,
solid line - $\alpha_0=\pi/3$, dotted line - $\alpha_0=0$.
Insets on (a), (c): cuts and integral paths on the $\omega$-plane;
inset of (b): $\varphi_{\bf k}^{\cal C} (t)$ for weak anisotropic perturbation;
inset of (d): example of temporal pattern for $\omega_{pl} \approx 2 \omega_{vK}$.
}\label{fig-1}
\end{figure}

Interestingly, the ratio of the magnitudes of the two components depends essentially
on the wavevector. Indeed, for $k \ll K$, and $\textstyle{t \rightarrow 0}$ we obtain the estimate
$\varphi_{k}^{\cal C} \approx k/2 K \ll 1$. From Fig. 1(b) we see that
at $k = K$  $\varphi_{k}^{\cal C} \approx 0.33$.
In the opposite case, $k \gg K$, we obtain $\varphi_{k}^{C} \approx 1$.
That is, the regular plasmon component is preferable excited by
long-wavelength perturbations.

Then, consider an initial perturbation of the anisotropic form
$\textstyle{\delta {\cal F}_{\bf k} ({\bf p}) =
\delta {\cal F}_k ( p) {\bf p n}/p}$,
with $\bf n$ being a unit vector of a preferential direction.
We denote the angle between
$\bf n$ and $\bf k$ as $\alpha_0$. For this case,  at $t=0$
the perturbed electron gas receives additional momentum and
nonzero velocity, while the electron density is unperturbed.
The calculation gives
\begin{equation}  \label{N-2}
N(\omega,{\bf k}) = i \frac{ \Phi_k^{0} \cos \alpha_0}{v_F k}
\left[\frac{\omega}{\sqrt{\omega^2 - v_F^2 k^2}} -1 \right]\,,
\end{equation}
where $\Phi_k^{0} $ formally is given by the second relationship
of Eq.~(\ref{N-1}). Calculating the time-dependent potential $\Phi_{\bf k} (t)$ with this
$N(\omega, {\bf k})$ function, one can use the analytical continuation discussed above,
as well as Eqs.~(\ref{phi-0}) - (\ref{phi-4}), and  the integration contour shown
in the inset of Fig.~1(a). The residue contribution to the potential is
\begin{equation} \label{N2-R}
\!\!\frac{\Phi_{\bf k}^R (t)}{ \Phi_{k}^0} =\! - 2 i \cos \alpha_0
\frac{[\omega_{pl}^2 (k) - v_F^2 k^2 ]^{3/2}}{v_F^3 k K^2} \sin[\omega_{pl} (k)\, t]\,.
\end{equation}
For the contour integral of Eq.~(\ref{phi-4}), we present the results
obtained for two limiting cases:
\begin{equation}  \label{PhiC-2}
\frac{\Phi^{\cal C}_{\bf k} (t)}{\Phi_{k}^0}  \!\approx\! i \cos \alpha_0 \!\!
\begin{cases}
\displaystyle{\left(\frac{k}{K}\right)^{\!2}
\left[\!\frac{J_0[v_F k t]}{v_F k t}\! -\!
 2 \frac{J_1[v_F k t]}{[v_F k t]^2}\right]\!, k\ll K,}\\
 \displaystyle{J_1 (v_F kt),
 \phantom{emptyemptyempty!!!!}k \gg K.}
\end{cases}
\end{equation}
At $t=0$, both contributions $\Phi_{\bf k}^R$ and $\Phi^{\cal C}_{\bf k}$ equal zero.
At $ t \gg 1/v_F k$, we obtain:
\begin{equation}
\frac{\!\Phi^{\cal C}_{\bf k} (t)}{ \Phi_{k}^0}  \!\approx\! i \sqrt{\frac{2}{\pi}}\cos \alpha_0 \!
\begin{cases}
\displaystyle{\left(\frac{k}{K}\right)^{\!2}\!
\frac{\cos{[v_F kt - \pi/4 ]}}{(v_F kt)^{3/2}}\,,}\!\!&k\ll K,\\
\displaystyle{\,\,\,\,\frac{\cos{[v_F kt -3 \pi/4]}}{(v_F kt)^{1/2}}\,,}&  k \gg K\,.
\end{cases}
\end{equation}
Thus, weak anisotropic initial perturbation also produces two component
oscillations with purely imaginary magnitude, $\Phi_{\bf k}^{R}$ and $\Phi_{\bf k}^{\cal C}$, with frequencies
$\omega = \omega_{pl} (k)$ and $\omega = v_F k$, respectively. The latter
are decaying in time. At small $t \approx 1/ v_F k$, these oscillations
are quite different from those discussed above, since such perturbation
does not generate directly a space charge and an electrostatic potential.
Both are developing with time.
The normalized potential $\textstyle{\varphi_{\bf k}^{\cal C} (t) = \Phi_{\bf k}^{\cal C}/\Phi_{k}^0}$
of these oscillations is illustrated in the inset of Fig.~1(b).
The oscillations magnitude depends on the angle $\alpha_0$.
The magnitude reach maxima for $\alpha_0 = 0,\,\pi$.
If $\alpha =\pi/2$, the perturbation does not excite the oscillations.

Finally, consider the extremely anisotropic initial perturbation:
$\delta {\cal F}_{\bf k} ({\bf p}) = 2\pi\delta {\cal F}_{k} ({p})\,
\delta[\alpha-\alpha_0]$, where $\alpha$ is the angle between ${\bf p}$ and
$\bf k$, $\alpha_0$ determines the anisotropy direction.
For the function $N(\omega,{\bf k})$ we obtain
\begin{equation} \label{N-3}
\displaystyle{N(\omega,{\bf k}) = \frac{i \Phi_k^{0}}{\omega -v_F k \cos \alpha_0 }},
\end{equation}
with $\Phi_{k}^0$ still defined by Eq.~(\ref{N-1}). The function (\ref{N-3})
has a pole at the $\omega'$-axis: $\omega = \omega_{vK}  ({\bf k}) \equiv v_F k \cos \alpha_0\,$.
That implies that the accepted above way of the analytical continuation
of the function $\Pi (\omega,  k)$ is no longer applicable.
Instead, one can use two cuts along semi-infinite lines in the lower part
of the $\omega$-plane: $\omega = \pm v_F k + i \omega^{''},\,\omega^{''} \leq 0$,
as shown in the inset of Fig.~1(c). Now $\Pi (\omega,  k)$ is continuous
across  the real axis everywhere excluding points $\omega = \pm v_F k$ and
undergoes jumps at the semi-infinite cuts.
The time-dependent total potential (\ref{phi-2}) consists of four contributions:
\begin{equation} \label{phi-0b}
\Phi_{\bf k} (t) = \Phi^{R}_{\bf k} (t) + \Phi^{{\cal C}_l}_{\bf k} (t)+ \Phi^{{\cal C}_r}_{\bf k} (t)+ \Phi^{vK}_{\bf k} (t)\,.
\end{equation}
In contrast to the previous cases, every contribution
has nonzero real and imaginary parts.
The term $\Phi^{R}_{\bf k} (t)$ is the plasmon contribution determined by
Eq.~(\ref{phi-3}) with $N(\omega, {\bf k})$ given by (\ref{N-3}),
\begin{eqnarray} \label{StrongPlasmons}
\displaystyle{\frac{\Phi^R_{\bf k} (t)}{\Phi_{k}^0} =
\frac{2 \left[\omega_{pl}^2 -v_F^2 k^2 \right]^{3/2}}{v_F^2kK(\omega_{pl}^2-v_F^2 k^2\cos^2{\alpha_0})}\times}\\ \nonumber
\displaystyle{\times\{\omega_{pl}\cos[\omega_{pl} t]-i v_F k\cos{\alpha_0}\sin{[\omega_{pl}t]}\}.}
\end{eqnarray}
The terms $\Phi^{{\cal C}_l}_{\bf k} (t)$ and $\Phi^{{\cal C}_r}_{\bf k} (t)$, are
the contour integrals around the left and right cuts,
respectively (see Fig.~1(c)). Main panels of Figs.~1(c) and (d) display real parts of the total contribution
$\textstyle{\Phi_{\bf k}^{\cal C}=\Phi_{\bf k}^{{\cal C}_l}+\Phi_{\bf k}^{{\cal C}_r}}$ from the contour integrals. Latter can be found
explicitly in the limiting case $t \gg 1 / (v_F \, k\,\sin^2{\alpha_0})$:
\begin{equation} \label{phicstrong}
\!\!\frac{\Phi^{\cal C}_{\bf k} (t) }{\Phi^0_{k}} \!\approx\!\frac{\sqrt{2/\pi}}{\sin^2{\!\alpha_0}}\!
\begin{cases}
\displaystyle{\!\!\left.\frac{k}{K}\right\{\cos{[v_F kt - 3\pi/4 ]}-i\cos{\alpha_0}\times}\\
\displaystyle{\left.\left.\phantom{\frac{k}{K}}\!\!\!\!\!\!\!\!\!\!\!\!\times\!\sin{\![v_F kt\!-\!3\pi/4]}\right\}\!\!\right/\!\!(v_Fk t)^{3/2}\!, k\!\ll\! K,}\cr
\displaystyle{\!\!\left.\frac{K}{k}\!\right\{\!\cos{[v_F k t \!-\! \pi/4 ]}-i\cos{\alpha_0}\times}\\
\displaystyle{\left.\left.\phantom{\frac{k}{K}}\!\!\!\!\!\!\!\!\!\!\!\times\!\sin{[v_F kt\!-\!\pi/4]}\right\}\!\!\right/\!\!(v_Fk t)^{1/2}, k\!\gg\!K.}
\end{cases}
\end{equation}
The last term in (\ref{phi-0b}) is generated by the additional pole at
$\omega = \omega_{vK} (\bf k)$:
\begin{equation} \label{VKPhi}
\displaystyle{\Phi_{\bf k}^{vK} (t) = \frac{\Phi_{k}^0\,\exp{[-i \omega_{vK} ({\bf k}) t]}}{1+K(i\cos{\alpha_0}/|\sin{\alpha_0}|+1)/k},}
\end{equation}
Thus, the strongly anisotropic initial perturbation produces four components
of the oscillations. As in previous cases, one of these components, $\Phi_{\bf k}^{R} (t)$,
is the sustained plasmon oscillations. Two other components,
$\Phi_{\bf k}^{{\cal C}_l} (t)$  and $\Phi_{\bf k}^{{\cal C}_r} (t)$, are oscillations
with 'carrier frequency' $\omega = v_F k$, they have different amplitudes and
time dependences at finite $t$ and both are similarly decaying in time.
Restoring the space-dependent factor $exp[i {\bf k r}]$, one can see that $\Phi_{\bf k}^{R}$ and $\Phi_{\bf k}^{\cal C}$ components correspond to {\em traveling} waves.

Additional fourth component, $\Phi_{\bf k}^{vK} (t)$, corresponds to
{\it undamped oscillations} with the frequency $\omega_{vK} ({\bf k})$.
This wave traveling along the wavevector $\bf k$ (or in opposite direction) can be interpreted as a
particle beam modulated in time and space and moving with the velocity
$v_F \cos{\alpha_0}$. In physics of three-dimensional plasma with the
regular dispersion of the electrons, such type of waves, with real
$\omega$ and $ k$, are known as the van Kampen modes~\cite{VKampen, Kadomtsev}.
Particularly, van Kampen has shown~\cite{VKampen}
that sharply anisotropic initial disturbance of the plasma generates waves
among which there is such a mono-energetic modulated wave.  To distinguish
the solution (\ref{VKPhi}) from other components of oscillations,
we will designate it as {\it a van~Kampen mode}.

At a given $k$, the van Kampen mode depends on two parameters, $\Phi_{k}^0$
and $\alpha_0$, which characterize the initial perturbation. When the angle of
the anisotropy direction, $\alpha_0$, changes from $\pi/2$ to $0$, the frequency
of this mode increases from 0 to $ v_F k$. At that
the magnitude of the wave varies from infinity ($\alpha_0=\pi/2$) to
zero ($\textstyle{\alpha_0 \rightarrow 0}$). In the latter case, when the pole $\textstyle{\omega_{vK} \rightarrow v_F k}$,
the $\Phi^{\cal C}$-contribution sharply increases as illustrated by numerical calculations presented in Figs.~1(c), (d).
Interestingly, in the Dirac plasma, the modulated electron beam corresponding to
the van Kampen mode is composed by {\it unidirectional} but not necessary
mono-energetic electrons, as it is seen from the accepted above form of the
initial perturbation of the electron distribution.

Note, the region on the $\omega$-$k$ plane, where the van Kampen modes can be
excited,  coincides formally with the region, for which the quantum theory
predicts the existence of the electron-hole pairs at low temperature~\cite{Castro-rev, PlasmonReview, BasovReview, PlasmonReview1, PlasmonReview2, PlasmonReviewTHz, Wunsch, Sarma}. As contrasted with such one-particle excitations, the van Kampen modes
are collective charge excitations.

At a given $\bf k$, all obtained oscillating components are spatially coherent.
Their superposition (\ref{phi-0b}) may demonstrate a complex temporal behavior,
for example, a beating effect between the plasmon waves and the van Kampen modes,
as shown in the inset of Fig. 1(d).
To characterize a complex temporal signal, it is useful to apply
the Fourier analysis. We performed such an analysis for obtained
oscillations (\ref{phi-0b}) and found the amplitude and phase of this signal
as functions of the frequency.
The amplitude shows very narrow peaks at frequencies
of the regular plasmons and the van Kampen modes, while respective phases
behave with sharp jumps, as it is typical near resonance.
At the frequency $\omega=v_F k$, corresponding to the
$\Phi_{\bf k}^{\cal C} (t)$  component of
oscillations, the amplitude
has a smeared spike, however the phase appears clear characteristic kink.
These features of the amplitude and phase of the collective excitations can be
verified by contemporary measurement techniques.

{\em In conclusion}, we have used the semiclassical approach to obtain transparent results
on the dynamics of collective excitations in Dirac 2D electron gas. By solving the initial value problem for
the system of Boltzmann-Vlasov and Poisson  equations with different forms of
initial disturbances of the distribution function and charge, we found that
collective excitations of the electrons with the Dirac spectrum are composed by
a few distinct components of the oscillations. Among these, there are always well known
{\em sustained plasmon oscillations} with the frequency given by Eq.~(\ref{zeros-1}).
Also, there exist \emph{new type} of oscillations with the carrier frequency $\omega = v_F k$;
they are of a transient character, decaying in time according to a
\emph{power law}. Mathematically, the latter component of the oscillations arises
due to the \emph{branching} feature of the polarizability function. Finally, a strongly anisotropic
initial disturbance generates {\em another component of
undamped oscillations} of frequency $\textstyle{\omega = v_Fk \cos{\alpha_0}}$, with $\alpha_0$
being the angle between the wavevector $\bf k$ and the anisotropy direction. We interpreted
these undamped oscillations in the form of an electron {\em unidirectional} beam and attendant
electrostatic potential both modulated in time and space, as van Kampen's
mode~\cite{VKampen} in the plasma of the Dirac electrons.

The results obtained demonstrate that the semiclassical approach
is adequate to describe the dynamics of collective excitations of THz spectral diapason
in the Dirac electron gas.


\begin{references}

\bibitem{Castro-rev}
A. H. Castro Neto, F. Guinea, N. M. R. Peres, K. S. Novoselov, and A. K.
Geim, Rev. Mod. Phys. {\bf 81}, 109 (2009).

\bibitem{SarmaReview}
S. Das Sarma, S. Adam, E. H. Hwang, and E. Rossi, Rev. Mod. Phys. \textbf{83}, 407 (2011).

\bibitem{NovoselovReview}
A. N. Grigorenko,	M. Polini	and K. S. Novoselov, Nature Photonics \textbf{6}, 749 (2012).

\bibitem{PlasmonReview}
X. Luo, T. Qiu, W. Lu and Z. Ni, Mater. Sci. Eng. R \textbf{74}, 351 (2013).

\bibitem{BasovReview}
D. N. Basov, M. M. Fogler, A. Lanzara, Feng Wang, and Yuanbo Zhang, Rev. Mod. Phys. \textbf{86}, 959  (2014).

\bibitem{PlasmonReview1}
F. J. Garcia de Abajo,  ACS Photonics, \textbf{1}, 135 (2014).

\bibitem{PlasmonReview2}
A. Politano, G. Chiarello, Nanoscale \textbf{6}, 10927 (2014).

\bibitem{PlasmonReviewTHz}
T. Low and P. Avouris, ACS Nano \textbf{8}, 1086 (2014).


\bibitem{Wunsch}
B. Wunsch, T. Stauber, F. Sols, and F. Guinea, New J. Phys. {\bf 8}, 318 (2006).

\bibitem{Sarma}
E. H. Hwang and S. Das Sarma, Phys. Rev. B {\bf 75}, 205418 (2007).

\bibitem{Ryzhii_gate}
V. Ryzhii, Jpn. J. Appl. Phys., \textbf{45}, 923 (2006).

\bibitem{NS-1}
R. Hillenbrand, T. Taubner,  F. Keilmann, Nature {\bf 418}, 159 (2002).

\bibitem{NS-2}
A. Huber, N. Ocelic, D. Kazantsev,  R. Hillenbrand, Appl. Phys. Lett. {\bf 87}, 081103
(2005).

\bibitem{Fei-1}
Z. Fei, G. O. Andreev, W. Bao, L. M. Zhang, A. S. McLeod, C. Wang, M. K. Stewart, Z. Zhao, G. Dominguez, M. Thiemens, M. M. Fogler, M. J. Tauber, A. H. Castro-Neto, C. N. Lau, F. Keilmann, and D. N. Basov, Nano Lett. {\bf 11}, 4701  (2011).

\bibitem{Chen-1}
J. Chen, M. Badioli, P. Alonso-Gonzalez, S. Thongrattanasiri, F. Huth, J. Osmond, M. Spasenovic, A. Centeno,
A. Pesquera, P. Godignon, A. Zurutuza Elorza, N. Camara, F. J. Garcia de Abajo, R. Hillenbrand, F. H. L. Koppens,  Nature {\bf 487}, 77 (2012).

\bibitem{Fei-2}
Z. Fei, A. S. Rodin, G. O. Andreev, W. Bao, A. S. McLeod, M. Wagner, L. M. Zhang, Z. Zhao, G. Dominguez, M. Thiemens, M. M. Fogler, A. H. Castro-Neto, C. N. Lau, F. Keilmann, D. N. Basov, Nature {\bf 487}, 82 (2012).

\bibitem{Fei-3}
Z. Fei, A. S. Rodin, W. Gannett, S. Dai, W. Regan, M. Wagner, M. K. Kiu, A. S. McLeod, G. Dominguez, M. Thiemens, M. M. Fogler, A. H. Castro-Neto, F. Keilmann, A. Zettl, R. Hillenbrand, M. M. Fogler, D. N. Basov, Nature Nanotechnology \textbf{8},  821 (2013).

\bibitem{Fei-4}
 M. Wagner, Z. Fei, A. S. McLeod, A. S. Rodin, W. Bao, E. G. Iwinski, Z. Zhao, M. D. Goldflam, M. K. Liu, G. Dominguez, M. Thiemens, M. M. Fogler, A. H. Castro-Neto, C. N. Lau, S. Amarie, F. Keilmann, D. N. Basov, Nano Lett. {\bf 2}, 894  (2014).


\bibitem{PRL-phase}
J. A. Gerber, S. Berweger, B. T. O’Callahan, and M. B. Raschke, Phys. Rev. Lett.
{\bf 113}, 055502 (2014).


\bibitem{Alonso}
P. Alonso-Gonzalez, A. Y. Nikitin, F. Golmar, A. Centeno, A. Pesquera, S. Velez, J. Chen, G. Navickaite, F. Koppens, A.
Zurutuza, F. Casanova, L. E. Hueso, R. Hillenbrand, Science, {\bf 344}, 1369 (2014).

\bibitem{Japan-1}
N. Kumada, S. Tanabe, H. Hibino, H. Kamata, M. Hashisaka, K. Muraki, T. Fujisawa, Nature Comm. {\bf 4}, 1363 (2013).

\bibitem{Japan-2}
N. Kumada, R. Dubourget, K. Sasaki, S. Tanabe, H. Hibino, H. Kamata, M. Hashisaka, K. Muraki and T. Fujisawa, New J. Phys. \textbf{16},  063055, (2014).


\bibitem{Mikhailov}
S. A. Mikhailov and K. Ziegler, Phys. Rev. Lett. \textbf{99}, 016803 (2007).

\bibitem{Landau-1}
L. D. Landau, J. Phys. (USSR) {\bf 10}, 25 (1946).

\bibitem{Principi}
A. Principi, G. Vignale, M. Carrega and M. Polini, Phys. Rev. B {\bf 88},
195405 (2013).


\bibitem{Ferro}
F. Chen, J. Xia, D. K. Ferry, and N. Tao, Nano Lett. {\bf 9}, 2571 (2009).


\bibitem{VKampen}
N. G. van Kampen, Physica {\bf 21}, 949 (1955).

\bibitem{Kadomtsev}
B. B. Kadomtsev,  Sov. Phys. Uspekhi {\bf 11}, 328 (1968).



\end{references}
\end{document}